# Grid-Brick Event Processing Framework in GEPS

Antonio Amorim, Luis Pedro
*Faculdade de Ciencias, University of Lisbon, Portugal*

Han Fei,  Nuno Almeida, Paulo Trezentos
*ADETTI, Edificio ISCTE, University of Lisbon, Portugal*

Jaime E. Villate
*Department of Physics, School of Engineering, University of Porto*

Experiments like ATLAS at LHC involve a scale of computing and data management that greatly exceeds the capability of existing systems, making it necessary to resort to Grid-based Parallel Event Processing Systems (GEPS). Traditional Grid systems concentrate the data in central data servers which have to be accessed by many nodes each time an analysis or processing job starts. These systems require very powerful central data servers and make little use of the distributed disk space that is available in commodity computers. The Grid-Brick system, which is described in this paper, follows a different approach. The data storage is split among all grid nodes having each one a piece of the whole information. Users submit queries and the system will distribute the tasks through all the nodes and retrieve the result, merging them together in the Job Submit Server. The main advantage of using this system is the huge scalability it provides, while its biggest disadvantage appears in the case of failure of one of the nodes. A workaround for this problem involves data replication or backup.

## 1. INTRODUCTION

In many scientific disciplines, the need for Petabyte data storage, processing and transferring is emerging as a crucial problem and simultaneously large computing and storage facilities are always scarce resources. The storage and computing capabilities are often temporarily and geographically distributed unevenly, sometimes redundant in one place while almost non-existent in other. With the scientific and technical applications becoming more and more sophisticated, many researchers, working and living in different places, have to cooperate in the same re-search project and access distributed computing resources.

It is unlikely that conventional methods can meet the demands of providing and sharing these resources. A blueprint of computational grids leveled at addressing these difficulties has been proposed. [1]

The Grid [2] is super-computing net, which can connect distributed mainframe computers, super-computers, as well as large numbers of desktop computing devices into easy-to-use computing facilities.

### 1.1. The LHC Computing Problem

In the Large Hadron Collider (LHC) accelerator at CERN, there are 25 million collisions taking place per second. Each collision contains about 1 MB of information. One single collision is called an "event". Each event is recorded by surrounding particle detectors for later processing and filtering to select the physically interesting ones. At the end of the selection process, events are recorded at a typical rate of 100 Hz. Considering the data intensive aspect of event processing problems, computational grids are a possible solution.

## 2. RELATED WORK

One interesting development was the Gfarm (Grid Data Farm) project [3][4] at KEK (High Energy Accelerator Research Organization) and ICEPP (International Center for Particle Physics, the University of Tokyo). A large scale Gfarm file is divided into several fragments and distributed across the disks in the Gfarm file system. A Gfarm file is a logical aggregation of physical file fragments distributed over many CPU nodes. When a job is submitted into the Gfarm server, it is redistributed to nodes, which contain the fragment database files. When the job is finished, the results will be retrieved across the network.

Parallel ROOT (PROOF) [5] is another event processing system. The ROOT client session creates a master server on a remote cluster, and then the master server in turn creates slave servers on all the nodes in the cluster. All the slave servers execute the user job in parallel. The master server distributes the event data packets to every slave server, carefully adjusting the packet size such that the slower slave servers get smaller data packets than faster slave servers. PROOF uses a TChain object to provide a single logical view of many geographically distributed physical files. The master server keeps a list of all generated packets per slave, so in case a slave failed then remaining slaves can reprocess its packets.

## 3. TRADITIONAL APROACH

The GRID paradigm puts a strong emphasis on sharing computing and data resources over the global network. The usual approaches that are implementing the GRID like GLOBUS[13] or DATAGRID[14] focus on the copying and sharing of files as well as defining the site gateways to the job submission and file transferring facilities. The defined GRID interfaces address the traditional job submission use case where applications have well defined input and output streams and almost no interaction with the user except trough a job description file. At each site the data is copied from the data center to the processing CPU, it is used as input and dropped after the application finishes. On the other end, if the data has to be accessed trough the network opening a global URL, even the fastest global networks are a problem due to the large acknowledgment time necessarily associated with large physical distances.

One important by product of the previous architecture is the fact that the GRID enabled applications can share resources but do not become necessarily faster except if the user explicitly splits his computing job into many jobs and explicitly tunes the application input for this task. The traditional approaches actually assume that the parallelization can be done at a second step inside each GRID center.

One can perhaps claim that this is not the most natural way in which computing applications tend to evolve. Interactivity and access to database servers has already became crucial and the GRID should attempt to deliver results to the user much faster by assuming that the parallelism over independent events that are processed in parallel. The parallelism over independent high-energy physics collisions must be exploited to define GRID services with the granularity of one event.

## 4.  GEPS APROACH

In GEPS system we make use of the Grid infrastructure, a back-end database, LDAP directory query, and PHP script web interface. The scalability of GEPS can be easily obtained through freely adding into or removing any grid computing and storage node. The main philosophy of GEPS is that it works like a portal. Behind the friendly appearance of GEPS, many Grid related details are hidden. Geographically distributed physicists can easily cooperate over the same event processing project, share dispersed events data file, stage jobs, query job status, share computing resources, transfer data file, and visualize events filtering results.

The main philosophy in which GEPS is based is that the data should not be moved when applying for a job submission. Data should be already distributed between the different Grid nodes. This is very a important issue if we just imagine the amount of data transfer that will be needed for processing tremendous amounts of jobs simultaneously, each one of them in the order of the Mbytes. The important issue of scalability can be easily satisfied with this approach because it's just a matter of adding more Grid nodes to the system which are very standard machines.

### 4.1.  The Events Application

Event processing application is programmed in C++ using the Root Toolkits [6]. Root is an object-oriented framework, aimed at solving the data analysis challenges of the high-energy physics discipline. It provides a large collection of specific utilities to manage information in an efficient way. Root provides not only an application programming interface (API), but an integrated Root tree class data file visualization environment. The creation of the Root data file has several steps. The first step is to create a structure to store all the raw information of the events - this process consists of the creation of a shared library, which contains all the variables of the event, track, vertices, as well as relation objects.

The next step is to create a Root tree for the storage of all the objects presented in the raw information file - the Root tree class is optimized to reduce storage space usage and enhance accession speed. Inside the Root tree there is one branch with all events, inside this branch are all event variables that include the tracks, vertices, and relations.

After data storage in the Root tree is completed, now it is the time for scrutinizing, one by one, which event will be the candidate that meets the processing standard. The calibration procedure based on the processing standard will be done on each event, and then the result will be stored in a new tree with the same structure.

Based on the Grid-enabled computing net, we can split event raw data into different parts. These can be stored in geographically distributed Grid environment, namely the resource nodes available for data processing. After that, we can stage processing and filtering procedures in a parallel mode, monitor the running application, collect results, merge the different results data into final data file, and retrieve/display the final data.

### 4.2.  General architecture

Figure 1 describes the GEPS architecture. GEPS provides a user friendly interface that is PHP based. This interface ends up being a world wide interface though which the user can easily access to the GEPS system. After entering in the main-page, a set of options are provided, each one of them associated with a different function in the GEPS system. The options are: Submit jobs, Retrieve information about a particular Grid node; get jobs status details. These options available though the web interface will be described in more detail in section 5.

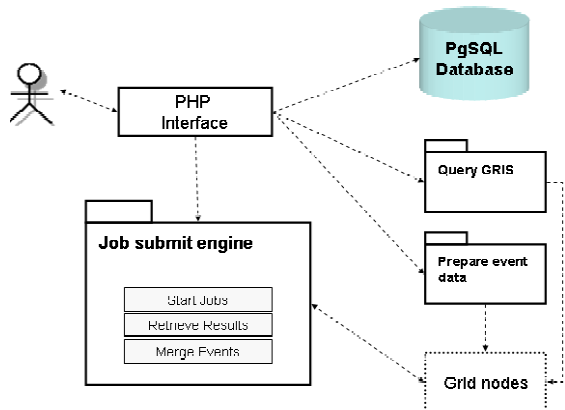

Fig 1 – GEPS general architecture. The PHP interface hides the implementation details from the user.

Whenever a user submits a job to the GEPS system, some information will be sent to the Meta-data catalogue and other to the Job Submission Engine (JSE). The Meta-data catalogue is a database that uses PostgreSQL (PgSQL) as Database Management System (DBMS). The JSE, through its *broker* that searches from time to time into the Meta-data catalogue, receives the information that a new job has been submitted for processing. JSE then sends the information to the Grid nodes that after the job has been processed it sends the result back to the JSE. Back there, the final result is merged from the various results coming from the different Grid nodes. At this time, JSE also updates the information in the Meta-data catalog. After this process the user is ready to retrieve the result of the processed jobs. Figure 2 shows schematically the dataflow between the different elements present in the GEPS system.

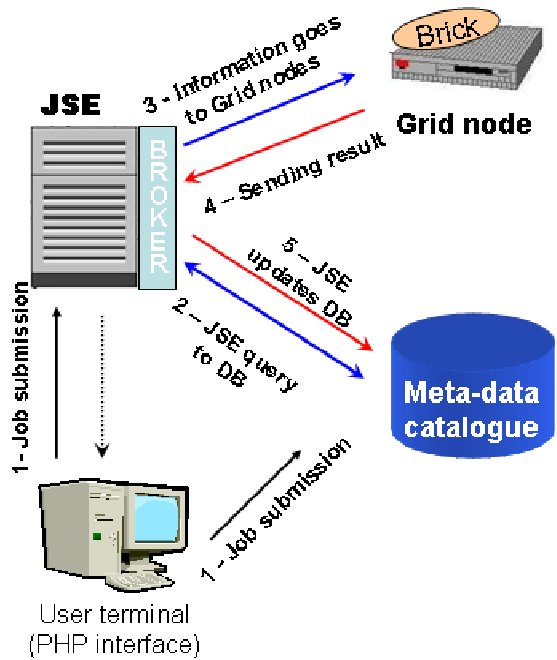

Fig 2 - GEPS data flow

For the information retrieval concerning the job status, the GEPS uses functions that queries the GRIS LDAP server.

The Grid JSE will parse the job specification tuple in the PgSQL database, analyze the job executing environment and raw event data distribution demands, synthesize the RSL sentences, submit the jobs and monitor the status of the submitted job.

### 4.3. GEPS Prototype Implementation

The GEPS prototype uses different technologies in order to accomplish the behavior that is expected. The technologies involved are:

- **Globus toolkit** – toolkit that provide GRID API functions
- **PgSQL** – for the Database Management System (DBMS) of the Meta data catalogue and to store other user related information;
- **LDAP** -  Query Grid node information
- **PHP** – For the web interface

The Globus Grid toolkit evolved out of the I-WAY high performance distributed computing experiment [7]. Before the Globus Grid became the de facto high performance computing environment, there were other candidate grid architectures, include using object-based technology and web technology [8].

Table 1: Globus components in GEPS

| Component | Usage |
|-----------|-------|
| GRAM | Executable staging |
| GRIS in MDS | Query Grid node information |
| GASS | Transfer raw data, retrieve remote results |

Table 1 lists the grid components used in GEPS. In the Grid job submission engine, the new job specification tuples are selected from the backend PgSQL database. For each new job, by parsing the job specification tuple, a job Resource Specification Language (RSL) sentence is formulated. After that, a raw data file is transferred (by using GASS components) in accordance with the setting of relevant resources and then, the GRAM component (globus-gram-client) is used for remotely submitting and managing job. In run time, *stdout* and *stderr* is defined in the RLS sentence. After all submitted jobs having finished, GASS file access functions are used for retrieving distributed event results.

The Globus Toolkit provides an information Monitoring and Discovery Service (MDS)[9], which acts as a resource information registry and discovery agent. The MDS includes a standard, configurable information provider framework called a Grid Resource Information Service (GRIS). GRIS is implemented as an OpenLDAP [10][11] server. Each Grid node can run a local GRIS.

Through GEPS, the end user can query properties of the grid nodes, *e.g.* how many processors are available at this moment, what bandwidth is provided, among others. The MDS provides two interfaces: interactive and programmatic. By default, a GRIS service is automatically configured and assigned to work on port 2135. In our GEPS, the grid-info routine obtains the overall Grid node information by querying this port through the LDAP protocol. The PHP script will call the grid-info routine to get the results.

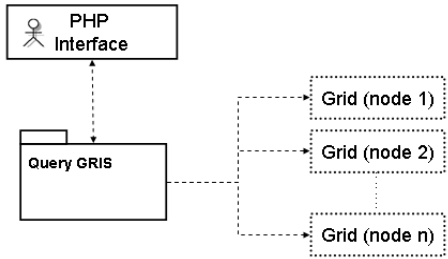

Fig 3 - Querying Grid node resource information through the LDAP protocol.

## 5. GEPS IN ACTION

In this section will give a very general overview to the options that the user disposes in the web interface to the GEPS system. In the main page (Fig. 3), users can select different options: Get GREED information; Submit a job, Retrieve jobs information.

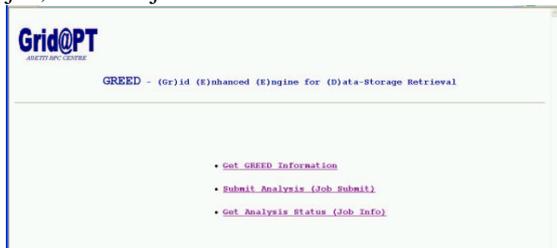

Fig 3 – GEPS main page.

For the job submission (Fig. 4) users need to fill in the web form according to their needs. The form includes options such as to which server will the job be submitted, a filter expression as well as a set of examples that can be used to help users to fill the filter expression field. After this process, the jobs will be sent to the grid nodes.

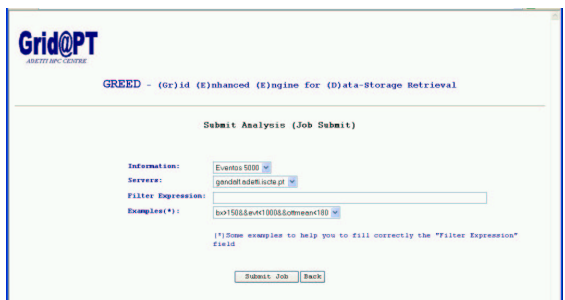

Fig 4 – GEPS – Submit a job to the system

THAT004

If the user wants to retrieve any information related to a particular Grid node it just need to select from which one in a form like the one in Figure 5.

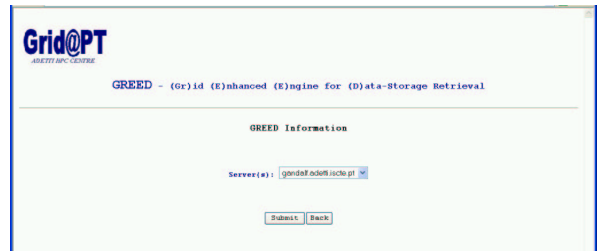

Fig 5 – Apply for retrieval of GREED information

A summary or the detailed information about the available Grid resources will be then presented.

The information about the job status it's presented like in Figure 6. This information includes:
- Which executable is being used;
- Where does the executable reside;
- To which Grid nodes is the executable going to be submitted;
- Where's the raw data file;

Fig 6 – GEPS – Job status information

## 6. TESTS AND RESULTS

From August to October in 2002, we tested 13 groups of raw event data, and with a total of 130 experiment executions (for decreasing the effect of system and network latency in executable staging and data transfer). The current GEPS demonstration prototype temporarily consists of two server, gandalf and hobbit. Because the GEPS topology structure has the feature of scalability, in the future more nodes can be easily incorporated. One of the advantages of computational grids is that any parts can be easily changed without any global effect.

Different granularities of event data will dramatically affect the overall performance of the GEPS system. This is reasonable, because with many smaller files of raw event data, the portion of system cost dedicated to raw data transfer will become larger in total execution time. Based on the event data file size, Figure 7 gives the relation between running only on hobbit and running in parallel between gandalf and hobbit. The unit on Y-axis is time cost in second, and the unit in X-axis is the number of events in raw event data file. In raw event file each event is about 1 MB in size. From the illustration we can easily see that the data file size of approximate 2000 events is a watershed. Data files consisting of less than 2000 events run in tightly coupled computing environments will have better performance. But usually our event raw data files can be easily much larger than 2000 events. From the results illustrated in Figure 7 we know that to some extent our GEPS has provided better performance.

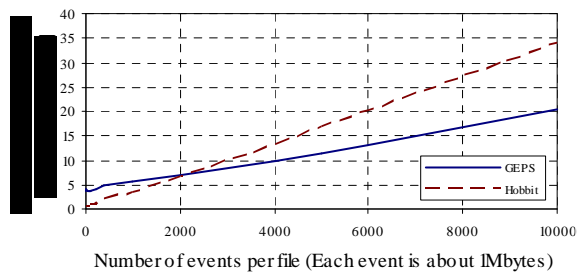

Fig 7 - Performance in GEPS and hobbit with different event raw data file sizes

The GEPS network connection is fast Ethernet. User defines raw event data distribution by using RSL sentence. Before a job can be submitted to grid gatekeeper through grid client API, raw event data will firstly be transferred to grid nodes in accordance with the raw event data distribution specification. GEPS currently uses globus gass file access API for transferring raw data and result file between gird nodes. Figure 6 only gives the comparison of processing time cost between GEPS and hobbit.

# 7. CONCLUSIONS AND FUTURE WORK

We have described the GEPS prototype, which provides an integrated Meta computing environment for event processing and filtering. In GEPS, Grid related details and relevant middleware specifics have been hidden from the end user. With GEPS the scalability of intensive event data storage becomes easier to achieve. Using GEPS, physicists in any place can easily administer and share distributed data and take advantage of distributed computing resources. This prototype has successfully incorporated to date innovative Grid concepts and mechanisms.

The smaller bandwidth and the larger latency due to the geographical distribution of the Grid computational resources are the main reason of parallel inefficiency. We are working on adding GridFTP into our prototype. Because multiple TCP streams and proper TCP buffer sizes are very important to obtaining better performance in TCP wide area links [12], we are trying to add this feature into the GEPS prototype. We are also exploring the feasibility of solving other physics problems in the GEPS prototype environment like:

- Error handling and fault-tolerance;
- Recover mechanisms for each node;
- Create a redundancy mechanism to recover from a malfunction in the nodes;
- Develop a storage mechanism to submit more work to the best nodes - Load balancing;

# 8. ACKNOWLEDGEMENTS

This work was supported by Fundação da Ciência e Técnologia under the grant CERN/P/FIS/43719/2001. The first author gratefully acknowledges the postdoctoral fellowship by the FCT. The third author would like to thank ADETTI (Associação para o Desenvolvimento das Telecomunicações e Técnicas de Informática) for their support to this work.

**THAT004**